\input{psfig}
\input epsf
\documentstyle[11pt]{article}

\advance\oddsidemargin by 8mm

\def\be{\begin{equation}}
\def\ee{\end{equation}}

\def\ztwo{{\cal Z}^{(2)}}

\def\half{\frac{1}{2}}
\def\third{\frac{1}{3}}

\def\ph3{$\phi^3$}
\def\gst{\gamma_{str}}
\def\sg{{s_{\scriptscriptstyle G}}}

\def\similar{\quad {\mathop\sim\limits_{\scriptscriptstyle
 N \to \infty} } \quad}

\def\gtwo{{{\cal G}^{(2)}(N)}}
\def\ggtwo{{\cal G}^{(2)}}

\def\tloop{{n_2^c}}
\def\mgap{\qquad \qquad}

\def\bG{{\bf G}}

\def\ltsim{\lower3pt\hbox{$\, \buildrel < \over \sim \, $}}
\def\gtsim{\lower3pt\hbox{$\, \buildrel > \over \sim \, $}}

\begin{document}
\psfull

%%%%%%%%%%%%%%%%%%%%%%%%%%%%%%%%%%%%%%%%%%%%%%%%%%%%%%%%%%%%%%%%%%%%%
%% The title and abstract
%%%%%%%%%%%%%%%%%%%%%%%%%%%%%%%%%%%%%%%%%%%%%%%%%%%%%%%%%%%%%%%%%%%%%
\vbox{\smash{\vbox{
\begin{flushright}
 \large OUTP 9503P \\
 \large 6th February 1995
\end{flushright}}}
\title{A Two Term Truncation of the Multiple Ising Model Coupled to 2d
Gravity}
\author{M. G. HARRIS \\ \\
  \small Theoretical Physics, University of Oxford,\\
  \small 1 Keble Road, Oxford OX1 3NP, UK \\
  \small E-mail address: harris@thphys.ox.ac.uk}
\date{}
\maketitle}
\begin{abstract}
We consider a model of $p$ independent Ising spins on a dynamical
planar \ph3 graph. Truncating the free energy to two terms yields an
exactly solvable model that has a third order phase transition from a
pure gravity region ($\gst=-1/2$) to a tree-like region ($\gst=1/2$),
with $\gst=1/3$ on the critical line. We are able to make an order of
magnitude estimate of the
value of $p$ above which there exists a branched polymer (ie
tree-like) phase in the full model, that is, $p \sim 13$--$23$, which
corresponds to a central charge of $c \sim 6$--$12$.

\end{abstract}

%%%%%%%%%%%%%%%%%%%%%%%%%%%%%%%%%%%%%%%%%%%%%%%%%%%%%%%%%%%%%%%%%%%%%
%% The text
%%%%%%%%%%%%%%%%%%%%%%%%%%%%%%%%%%%%%%%%%%%%%%%%%%%%%%%%%%%%%%%%%%%%%

\section{Introduction}

There has been much interest in studying models of conformal matter
coupled to 2d quantum gravity and in particular in investigating the
nature of the $c=1$ barrier, beyond which KPZ theory~\cite{KPZ} breaks
down. These models tend to exhibit a branched polymer phase for large
enough values of the central charge $c$, and our aim is to shed some
light on the extent of this phase for the multiple Ising model.

This letter extends the work of a previous paper~\cite{paper2} in
which we studied a model of $p$ independent
Ising spins coupled to 2d gravity, the
central charge of this model being $c=p/2$. The partition function of
the model is
\be
Z_N(p) = \sum_{G \in {\cal G}} \frac{1}{\sg} \left( Z_G \right)^p,
\ee
where the sum is over some set ${\cal G}$ of connected planar
$N$-vertex \ph3 graphs.
The symmetry factor $\sg$ is equal to the order of the symmetry
group of the graph $G$. $Z_G$ is the Ising partition function for a
fixed graph $G$ with a single spin per vertex and coupling constant
$\beta$, that is,
\be
Z_G = \frac{1}{Z_0}
\sum_{\{S\}} \exp \left( \beta \sum_{<i j>} S_i S_j \right) \ ,
\ee
with
\be
Z_0 = 2^N \left( \cosh \beta \right)^\frac{3N}{2}.
\ee
$S_i$ is the spin on the $i$-th vertex ($S_i = \pm 1$) and the sum in
the action is over nearest neighbours on the \ph3 graph.

As in our previous paper we define three different versions of the
model. For model~I the set ${\cal G}$ consists of all connected planar
$N$-vertex \ph3 graphs. In model~II the set ${\cal G}$ is restricted to
one-particle irreducible  graphs and for
model~III only two-particle irreducible graphs are used.

\begin{figure}[b]
\caption[l]{Tree-like graph}
\label{fig:tree}
\begin{picture}(100,35)(-45,0)
\epsfbox{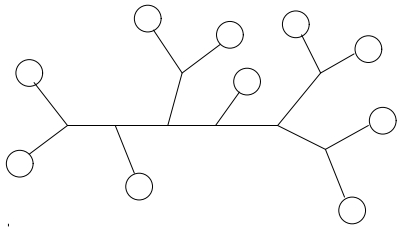}
\end{picture}
\end{figure}

It has been proven~\cite{paper2} that in the limit of large $p$
tree-like graphs dominate the partition function, for model~I (see
fig~\ref{fig:tree} for an example of a tree graph).
By solving a model in which the free energy of
the Ising partition function was truncated to a single term we managed
to show how this change to dominance occurs, for small $\beta$. In
this approximation the change was continuous and for finite $p$ there
was no phase transition to a tree-like region (which we would
associate with the branched polymer region occurring in other
similar models).

The purpose of this letter is to extend the calculation by including
two terms in the truncated free energy. We will show that for this
two-term truncation of model~I there exists a third order phase
transition from a pure gravity region (with $\gst=-1/2$) to a
tree-like region (for which $\gst=1/2$) and that $\gst=1/3$ on the
critical line. The location of this critical line approximates to that
in the full untruncated model and allows us to estimate the value of
$p$ for the point at which the tree-like, unmagnetized and magnetized
phases meet in this model; the result is $p^* \sim 13$--$23$,
corresponding to a central charge of $c \sim 6$--$12$.
In this letter we present the main details of the calculation, further
information about the solution can be found in
reference~\cite{thesis}.

\section{Definition of the two-term truncated model}

\begin{figure}[tbh]
\caption[l]{Cutting 1-loops}
\label{fig:cut}
\begin{picture}(100,20)(0,0)
\centerline{\epsfbox{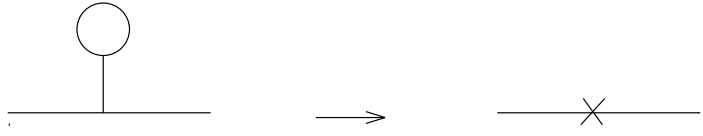}}
\end{picture}
\end{figure}

For a given \ph3 graph $G$, the Ising partition function can be
written in terms of $t \equiv \tanh \beta$ as a high temperature
expansion,
\be
Z_G = 1 + \sum_l n_l t^l,
\ee
where $n_l$ is the number of closed (but possibly disconnected) loops
in the graph, which contain $l$ links and use no link more than once.
Defining $\mu_G$ by
\be
\mu_G = \lim_{N \to \infty} \frac{1}{N}  \log Z_G ,
\ee
we have for model~I,
\be
 \mu_G =
\frac{1}{N} \left[ n_1 t + \left( n_2 - \half {n_1}^2\right) t^2 +
\cdots \right] .
\ee
Defining $\tloop$ to be the number of connected loops of length two
(hereafter referred to as ``2-loops'') then $n_2 = \tloop + \half n_1
(n_1 -1)$ and we have
\be
\mu_G =
\frac{1}{N} \left[ n_1 t + \left( \tloop - \half {n_1}\right) t^2 +
O(t^3) \right] .
\ee
Suppose that we truncate $\mu_G$ to the first two terms of the series,
denoting this new quantity by $\mu_G^T$ and consider the model with
the partition function
\be
\label{eq:partfunc}
Z_N(p) = \sum_{G \in {\cal G}} \frac{1}{\sg} e^{\mu_G^T N p},
\ee
which we will refer to as the two-term truncated model.
{}From now on ${\cal G}$ is taken to be the set of model~I graphs.
In the limit of small $\beta$ we might expect it to reproduce the
behaviour of the full model. For larger $\beta$ it will be a bad
approximation to the original model and in particular it has no magnetized
phase. However we can modify the truncated model slightly so that it
is a reasonable approximation to the full model for large $p$ at any
value of $\beta$. The key to doing this is to note that one can factor
out the entire contribution to $Z_G$ resulting from loops of length
one (``1-loops'') giving
\be
Z_G = (1+t)^{n_1} Z_G',
\ee
where $Z_G'$ is the partition function on the graph $G'$ for which
1-loops have been cut off (leaving vertices of coordination number
two, see fig~\ref{fig:cut}).
It is impossible to factorize all the contributions from
2-loops in the same fashion, but suppose that we consider the
partition function
\be
{Z_G^{''}} = (1+t)^{n_1} (1+t^2)^{\tloop} = 1 + \sum_l n_l' t^l.
\ee
This corresponds to a model in which we count $n_l'$ the number of
(possibly disconnected) loops of length $l$, but restrict each of the
connected parts to be of length two or less.
This is a good approximation to the full model if $t$ is small or if the
dominant graphs only have short loops in them (which is precisely what happens
for large $p$ --- since we know that tree-like graphs dominate in this
limit). Thus we might expect the modified two-term truncated model to
give a reasonably accurate indication of the transition to a tree-like
phase even for values of $\beta$ which are not small.
The corresponding $\mu_G^T$ is
\be
\label{eq:mugt}
\mu_G^T = \frac{1}{N} \left[ n_1 \log(1+t) + \tloop \log \left( 1 +
t^2 \right) \right].
\ee
In this letter we will solve exactly this modified truncated model,
defined by equations (\ref{eq:partfunc}) and (\ref{eq:mugt}), and
relate the results to the untruncated model. For technical reasons it
is much easier to solve the model if we use \ph3 graphs which have a
``root''; that is, in the terminology of $\lambda \phi^3$ theory, we
are going to use 1-legged Green functions rather than vacuum diagrams
- this will not change the free energy and should not affect any of
our conclusions. In this case we have
\begin{eqnarray}
Z_N(p) &=& \sum_{G \in {\cal G}} \exp p \left[ n_1 \log(1+t) + \tloop
\log \left(1 +t^2 \right) \right] \\
       &=& \sum_{n_1=0}^{\infty} \sum_{n_2=0}^{\infty} \
{\cal G}_r^{(1)}(N,n_1,\tloop) \ v^{n_1} y^{\tloop} ,
\end{eqnarray}
where we have defined
\be
v \equiv (1+t)^p \ , \ \ \  y \equiv \left( 1 + t^2 \right)^p
\ee
and ${\cal G}_r^{(1)}(N,n_1,\tloop)$ is the number of $N$-vertex
rooted model~I graphs with $n_1$ 1-loops and $\tloop$ 2-loops; the
subscript ``r'' indicates that we are using rooted graphs. The
grand canonical partition function is
\be
\label{eq:defcalzr}
{\cal Z}_r = \sum_{N=1 \atop odd}^{\infty} e^{- \mu N} Z_N(p) =
\sum_{N=1 \atop odd}^{\infty} \sum_{n_1=0}^{\infty}
\sum_{n_2=0}^{\infty} {\cal G}_r^{(1)}(N,n_1,\tloop) \
 x^N v^{n_1} y^{\tloop} ,
\ee
where we have put $x \equiv \exp (- \mu)$.

\section{One-term truncation of model~II}

In the calculation that follows we will need the generating function for
${\cal G}^{(2)}{(N,\tloop)}$, the number of $N$-vertex (unrooted)
model~II graphs
with $\tloop$ 2-loops. Determining this is equivalent to solving the
one-term truncation of model~II and the calculation follows that given
in~\cite{paper2} for models~I and~III. As in that paper one can derive
a recurrence relation valid for $N \ge 4$,
\be
\label{eq:rrtwo}
\left( \frac{3N}{2}-2 \tloop \right) \ \ggtwo(N,\tloop) + 2 (\tloop+1)
\ \ggtwo(N,\tloop+1) = (\tloop+1) \ \ggtwo(N+2,\tloop+1) .
\ee
The generating function is defined as
\be
{\ztwo}(x,y) = \sum_{N=4 \atop even}^\infty
\sum_{\tloop=0}^\infty \ggtwo(N,\tloop) \ x^N y^{\tloop} .
\ee
This is given by
\be
\label{eq:solntwo}
{\ztwo} = \frac{1}{6} x^2 \left( h^{-3} - 3 y + 2 \right) - \half
\log h + \sum_{N=4 \atop even}^{\infty} \gtwo \ x^N h^{-3N/2} ,
\ee
where $h \equiv 1 - x^2 (y-1)$ and $\gtwo$ is the number of $N$-vertex
model~II graphs,
\be
\label{eq:counttwo}
\gtwo = \frac{2^\frac{N}{2} \left( \frac{3N}{2} -1 \right)!}{
\left(\frac{N}{2}\right)! \left(N+2\right)!} \similar
e^{\half \log \left( \frac{27}{2} \right) N} N^{- \frac{7}{2}} .
\ee

\section{Cayley Trees}

In the previous section we gave the generating function for
$\ggtwo(N,\tloop)$, which is the number of
one-particle irreducible (denoted ``1PI'') graphs with a given number
of vertices and 2-loops. If we could generalize this generating
function to the case of 1PI graphs with $m+1$ legs, then connecting
together such graphs into Cayley trees would give model~I graphs and
allow us to keep track of both $n_1$ and $\tloop$.

\begin{figure}[hbt]
\caption{(a) 1PI Green function, $\bG_{1 m}^{'}$ (b) Generating function
for trees, $T$ (c) Example of a rooted tree}
\label{fig:roottree}
\begin{picture}(100,30)(0,0)
\centerline{
\epsfbox{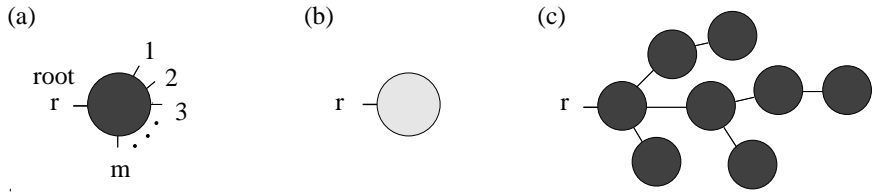}}
\end{picture}
\end{figure}
\begin{figure}[hb]
\caption{Equation for $T$}
\label{fig:eqntree}
\begin{picture}(100,25)(0,0)
\centerline{
\epsfbox{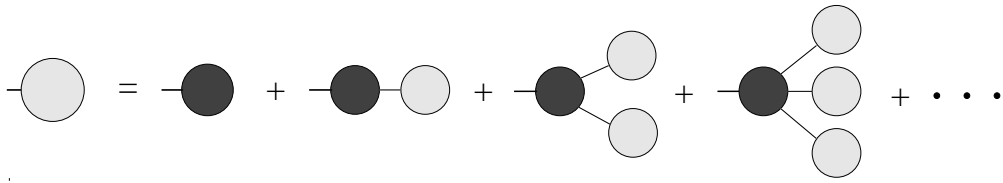}}
\end{picture}
\end{figure}

The generating function for 1PI graphs with a root and $m$ other legs
will be denoted by $\bG_{1m}^{'}$ and drawn as in
fig~\ref{fig:roottree}a (sometimes the root will be labelled with an
``r'' to distinguish it from the other legs).
It is a sum over such graphs weighted with
factors of $x^N y^{n_2^c}$. Note that although the legs have been
drawn on the exterior of the Green function some of them may in fact
be attached to links in the interior.

\begin{figure}[hbt]
\caption{(a) Example of a graph with two exits; (b) \& (c) Special case}
\label{fig:exit}
\begin{picture}(100,25)(0,0)
\centerline{\psfig{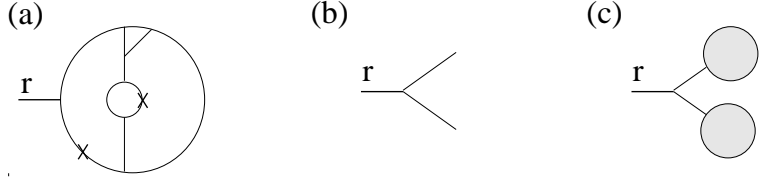}}
\end{picture}
\end{figure}

The generating function for trees will be
denoted $T$ and is drawn as in fig~\ref{fig:roottree}b.
Fig~\ref{fig:roottree}c gives an example of a tree contributing to
$T$. Figure~\ref{fig:eqntree} shows that $T$ satisfies the equation
\be
\label{eq:trelation}
T= \sum_{m=0}^\infty \bG_{1m}^{'} T^m .
\ee
It will be convenient to define a slightly different function $\bG_{1m}$
which generates 1PI graphs with a root and $m$ ``exits''. An exit is
defined to be
a place at which a tree can be hung (in either of two directions) and
is drawn as a link with a cross on it. For example fig~\ref{fig:exit}a
contributes to $\bG_{12}$ and there are four corresponding contributions
to $\bG_{12}^{'}$. In general $\bG_{1m}^{'} = 2^m
\bG_{1m}$, but
it should be noted that fig~\ref{fig:exit}b can not be represented in
this fashion and must be excluded from $\bG_{12}$. To compensate an
extra term (fig~\ref{fig:exit}c), $x T^2$, must be added to the equation,
giving
\be
\label{eq:treltwo}
T= \sum_{m=0}^{\infty} \bG_{1m} (2T)^m +x T^2 .
\ee
This formula then correctly distinguishes between graphs which differ
only in that a tree has been hung off a link in the opposite direction
(these correspond to different triangulated surfaces).

Now define a generating function for the $\bG_{1m}$,
\be
\label{eq:genexit}
\Gamma(x,y,z) = \sum_{N>0} \ \sum_{n_2^c=0}^{\infty} \ \sum_{m=0}^{\infty}
G_{1m}(N,n_2^c) \ x^N y^{n_2^c} z^m,
\ee
where $G_{1m}(N,n_2^c)$ is the number of 1PI graphs with one root, $m$
exits, $N$ vertices and $n_2^c$ 2-loops, that is,
\be
\bG_{1m}= \sum_{N>0} \sum_{n_2^c=0}^{\infty}
 G_{1m}(N,n_2^c) \ x^N y^{n_2^c} .
\ee
This is what is needed,
except for the fact that the 1-loops have been ignored. The only
contribution to (\ref{eq:genexit}) from 1-loops occurs in $\bG_{10}$ and
is equal to $x$. The 1-loops will be weighted with an extra factor of
$v$ and this is achieved by defining
\be
\label{eq:tildegamma}
{\tilde \Gamma}(x,y,z,v)= \Gamma(x,y,z) -x +xv.
\ee
Hence from (\ref{eq:treltwo}) we have that $T$ must satisfy
\be
\label{eq:maineqn}
T= {\tilde \Gamma}(x,y,2T,v) +x T^2
\ee
and then the solution $T(x,y,v)$ is equal to ${\cal Z}_r$ defined in
(\ref{eq:defcalzr}). The next task is to find the function ${\tilde
\Gamma}(x,y,z,v)$.

\section{Determination of ${\tilde \Gamma}(x,y,z,v)$}

In this section we are going to determine $\Gamma(x,y,z)$ the
generating function for $G_{1m}(N,n_2^c)$.
However it is useful to find $G_{10}(N,n_2^c)$ first, as this
will provide a boundary condition.
Below we show that
\be
\label{eq:gonezero}
G_{10}(N \! - \! 1,n_2^c \! - \! 1)= 2 n_2^c \ \ggtwo(N,n_2^c) ,
\ee
for $N \ge4$ ($N$ is even).

\begin{figure}[hbt]
\caption{(a) Rooting a model II graph (b) Special case}
\label{fig:rootmod2}
\begin{picture}(100,23)(0,0)
\centerline{
\epsfbox{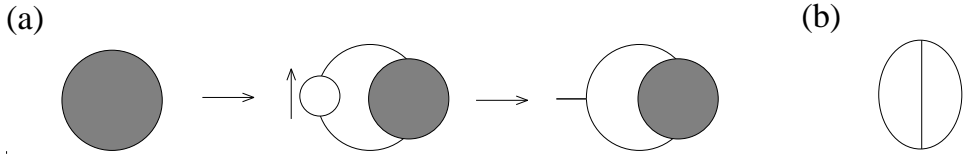}}
\end{picture}
\end{figure}

Take an $N$-vertex model~II graph $G$ with $n_2^c$ 2-loops. It has
$2\tloop$ orientated 2-loops (ie we count each 2-loop twice, once with
an arrow going in one direction and once with the arrow reversed).
These can be grouped into equivalence
classes of size $\sg$; each class consists of 2-loops which can be
transformed into each other under the action of the symmetry group.
Removing an orientated 2-loop and replacing it
with a leg (as shown in fig~\ref{fig:rootmod2}a) produces a graph with
$N-1$ vertices and $(\tloop -1)$ 2-loops. Note that we are ignoring the
special case fig~\ref{fig:rootmod2}b, for which $\tloop =3$ and $\Delta
\tloop = -3$; so the resulting formula will not apply for $N=2$. There
is a one-to-one correspondence between equivalence classes of
orientated 2-loops and rooted graphs. The number of equivalence
classes is
\be
\sum_{G:II \atop (N,\tloop)} \frac{2 \tloop}{\sg} = 2 \tloop \
\ggtwo(N,\tloop) ,
\ee
where the subscript on the summation means that we are summing over
model~II graphs labelled with a $G$, which have $N$ vertices and
$\tloop$ 2-loops. Equating this with the number of rooted graphs
$G_{10}(N \! - \! 1,\tloop \! - \! 1)$ gives the required result.

Now,
\be
\Gamma(x,y,z=0) = \sum_{N>0} \ \sum_{\tloop =0}^{\infty} G_{10}(N,\tloop)
\ x^N y^\tloop
\ee
and substituting (\ref{eq:gonezero}) into this yields
%\begin{eqnarray}
%\label{eq:boundary}
%\Gamma(x,y,z=0) &=& \frac{2}{x} \frac{\partial}{\partial y} \left[
%\sum_{N=4 \atop even}^{\infty} \sum_{\tloop =0}^{\infty}
%\ggtwo(N,\tloop) \ x^N y^\tloop \right] +x \\
% &=& \frac{2}{x} \frac{\partial {\cal Z}^{(2)}}{\partial y} +x ,
%\end{eqnarray}
\be
\label{eq:boundary}
\Gamma(x,y,z=0) =
 \frac{2}{x} \frac{\partial {\cal Z}^{(2)}}{\partial y} +x ,
\ee
where ${\cal Z}^{(2)}(x,y)$ is given by (\ref{eq:solntwo}).

\begin{figure}[hbt]
\caption{Adding an exit to: (a) an ordinary link (b) a 2-loop (c) a 1-loop}
\label{fig:addexit}
\begin{picture}(100,20)(0,0)
\centerline{\psfig{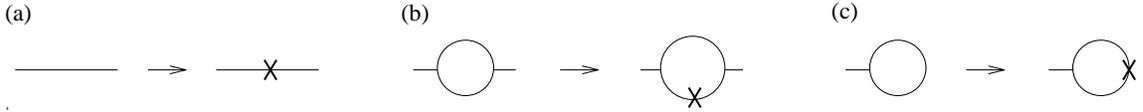}}
\end{picture}
\end{figure}
Having determined the number of rooted $N$-vertex 1PI graphs with
$\tloop$ 2-loops, we wish to generalize this to the case in which the
graphs have $m$ ``exits''.
Consider adding an exit to a graph which has $m-1$ exits and $N-1$
vertices; this will produce an $N$-vertex graph with $m$ exits. If the
exit is added to an ordinary link then $\tloop$ is unchanged
(fig~\ref{fig:addexit}a), if added to a 2-loop then $\Delta \tloop =-1$
(fig~\ref{fig:addexit}b) and if added to a 1-loop then $\Delta \tloop =
+ 1$ (fig~\ref{fig:addexit}c). Ignoring the special case
(fig~\ref{fig:addexit}c) for a moment, then we wish to make $N$-vertex
graphs, with $m$ exits and $\tloop$ 2-loops.
Bearing in mind that each such graph can be produced
in $m$ ways (ie any of the $m$ exits could be the one that we have
just added), we have
\be
\label{eq:exitrec}
m \ G_{1m}(N,\tloop) = L_1 G_{1 \ m-1}(N \! - \! 1,\tloop) +
L_2 G_{1 \ m-1}(N \! - \! 1,\tloop +1) ,
\ee
where $L_1$ is the number of ordinary links in a graph with $N-1$
vertices, $m-1$ exits and $\tloop$ 2-loops, and $L_2$ is the number of
links lying on a 2-loop in a graph with $\tloop +1$ such loops. One can
easily show that
\begin{eqnarray}
L_1 &=& \frac{3}{2} \left( N-(m+1) \right) + m - 2 \tloop ,\\
L_2 &=& 2(\tloop +1).
\end{eqnarray}
Thus we have a recurrence relation for the coefficients of
 $\bG_{1m}$ in terms of those in $\bG_{1 \ m-1}$
and a formula for $\bG_{10}$. Note that (\ref{eq:exitrec}) does not apply
for $N=2$, $m=1$ due to the special case (fig~\ref{fig:addexit}c).

Using (\ref{eq:exitrec}) one can derive the following differential
equation for $\Gamma(x,y,z)$,
\be
\frac{3}{2} x^2 \frac{\partial \Gamma}{\partial x} + 2x (1-y)
\frac{\partial \Gamma}{\partial y} - \left( \half x z +1 \right)
\frac{\partial \Gamma}{\partial z} - \half x \Gamma + x^2 (y-1) =0 .
\ee
The solution is straightforward and gives $\tilde \Gamma$  as
\be
\label{eq:thesolution}
\tilde \Gamma (x,y,z,v) = \sqrt{1-xz} \ \Gamma_0 \left( x
(1-xz)^{-\frac{3}{2}} , 1+ (y-1) (1-x z)^2 \right) + x^2 z (y-1) +
x(v-1) ,
\ee
where $\Gamma_0(x',y') \equiv \Gamma(x',y',z'=0)$.
Using (\ref{eq:boundary}) and (\ref{eq:solntwo}),
this is given by
\be
\label{eq:definegzero}
\Gamma_0(x',y') = x' \left[ \frac{x'^2}{h^4} + \frac{1}{h} +
\frac{3}{h} \sum_{N=4 \atop even}^\infty \gtwo \ x'^N N h^{-3N/2} \right],
\ee
with $h \equiv 1 -x'^2 (y'-1)$.

\section{Solution of the model}
Putting $z=2T \equiv 2 {\cal Z}_r$ and using (\ref{eq:maineqn})
gives
\be
\label{eq:zeqn}
xz = 2x \tilde \Gamma (x,y,z,v) + \half (xz)^2 .
\ee
As $x$ is increased from zero there is a non-analyticity at some
critical value, $x_c$. This will give us the free energy of the model
through $\mu_c = - \log x_c$. In order to determine $x_c$ we first
need to find an expression for the grand canonical partition function
${\cal Z}_r$.

 Defining $\epsilon  \equiv  y-1 $, $\eta  \equiv  v-1$ and
\begin{eqnarray}
Y & \equiv & \sqrt{1-xz} \\
Q & \equiv & \frac{x}{(1-xz)^{\frac{3}{2}}} = \frac{x}{Y^3} ,
\end{eqnarray}
we can eliminate $x$ and $z$ from (\ref{eq:zeqn}) in favour of $Y$ and
$Q$ giving
\be
\label{eq:chvarone}
\left(1-Y^2\right) = \half \left(1- Y^2 \right)^2 + 2 Y^3 Q \left[
Y \Gamma_0 \left(Q,1+\epsilon Y^4 \right) + Y^3 Q \left( 1-Y^2 \right)
\epsilon + Q Y^3 \eta \right] .
\ee
Now defining
\be
\label{eq:defineH}
H \equiv \frac{Q^\frac{2}{3}}{1- \epsilon Q^2 Y^4} =
\frac{x^\frac{2}{3}}{1-xz- \epsilon x^2} =
\frac{x^\frac{2}{3}}{Y^2 - \epsilon x^2} ,
\ee
we have using (\ref{eq:definegzero})
\be
\label{eq:defineF}
\Gamma_0 \left(Q,1+ \epsilon Y^4 \right) = Q^\third \left[
H^4 + H + 3H \sum_{N=4 \atop even} \gtwo N H^\frac{3N}{2} \right]
= Q^\third F(H) ,
\ee
where we have defined a function $F(H)$. Putting this into
(\ref{eq:chvarone}) gives
\be
\label{eq:chvartwo}
0 = Y^4 - 1 + 4 \left[Y Q^\third \right]^4 F(H) + 4 \left[ Y Q^\third
\right]^6 \left( \epsilon \left(1-Y^2\right) + \eta \right).
\ee
Noting that $Y Q^\third= x^\third$ we can use (\ref{eq:defineH}) to
eliminate $Y$ and write (\ref{eq:chvartwo}) in terms of $H$ and $f
\equiv x^\frac{2}{3}$,
\be
\label{eq:sixthorder}
\left( 3 \epsilon^2 \right) f^6 + \frac{2 \epsilon}{H} f^4 - 4 f^3
\left(\eta + \epsilon \right) - f^2 \left( 4 F(H) + \frac{1}{H^2} \right)
+1 =0 .
\ee
This is a sixth order polynomial in $f$, giving $f=f(H)$.

\begin{figure}[hbt]
\caption{(a) $\Gamma_2$ (b) $\Gamma_2^{-1}-1$}
\label{fig:selfenergy}
\begin{picture}(100,20)(0,-5)
\centerline{\psfig{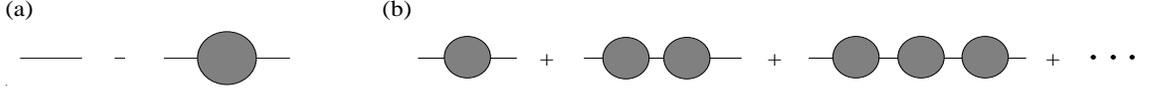}}
\end{picture}
\end{figure}

To proceed further we require a more compact form for $F(H)$.
Br\'ezin {\it et al}~\cite{BIPZ} give a formula for the 2-legged 1PI
generating function $\Gamma_2$ (equation 64 of that paper),
\be
\Gamma_2 = \frac{(1- 2 \tau )^2}{1-3 \tau} , \qquad {g'}^2= \tau (1- 2
\tau )^2
\ee
where each vertex has a weight of $g'$ (note that $g'$ is equivalent to
$3g$ in their notation).
 Now if we take each $N$-vertex model~II graph, pull
out an orientated link from each equivalence class (of size $\sg$) and
cut it, then we get
\be
\sum_{G:II \atop (N)} \frac{3N}{\sg} = 3N \gtwo
\ee
2-legged Green functions. Each possible connected 2-legged Green
function is generated exactly once, so that denoting the number of
such functions by $G_2(N)$ we have $G_2(N)= 3N \gtwo$.
Thus defining a new symbol $ \hat \Gamma$,
\be
\hat \Gamma \equiv \sum_{N=4 \atop even}^\infty 3N \gtwo {g'}^N =
\sum_{N=4 \atop even}^\infty G_2(N) {g'}^N .
\ee
However $\Gamma_2$ generates 1PI 2-legged Green functions and thus the
right-hand side of the above equation is given by $\Gamma_2^{-1} -1 -
{g'}^2$ (see fig~\ref{fig:selfenergy});
the ${g'}^2$ subtracts off the $N=2$ case.
The critical value of $g'$ is ${g'}_c= \sqrt{\frac{2}{27}}$ and the
corresponding critical value of $\tau$ is $\tau_c= 1/6$.
Hence,
\begin{eqnarray}
F(H) & = & H^4 + H + H \hat \Gamma (g'=H^{3/2}) \\
     & = & H \Gamma_{2}^{-1}(g'=H^{3/2}) .
\end{eqnarray}
We are lead to define the variable $\tau$ by
\be
H^3 = \tau (1-2 \tau)^2
\ee
and thus
\be
4 H^2 F(H) +1 = (1+6 \tau) (1- 2\tau) .
\ee
Now defining $k \equiv f \sqrt{H/\tau}$ we can rewrite
(\ref{eq:sixthorder}) as
\be
3 \epsilon^2 k^6 \tau^2 + 2 \epsilon k^4 \tau - 4 k^3 (\eta +
\epsilon) \tau (1-2 \tau) - k^2 (1+ 6 \tau) (1-2 \tau) + (1-2 \tau)^2
=0 .
\ee
This is a quadratic in $\tau$ and can be rewritten as
\be
\label{eq:tauquad}
A(k) \tau^2 + B(k) \tau + C(k) = 0
\ee
with
\begin{eqnarray}
\label{eq:funca}
A(k) &=& 3 \epsilon^2 k^6 + 8 k^3 (\eta+\epsilon) + 12 k^2 +4 \\
B(k) &=& 2 \epsilon k^4 - 4 k^3 (\eta +\epsilon) - 4 k^2 - 4 \\
\label{eq:funcc}
C(k) &=& 1- k^2 .
\end{eqnarray}
The aim of this calculation is to find the closest singularity to the
origin for the grand canonical partition function, ${\cal Z}_r$ (note
that ${\cal Z}_r = z/2$) and thus we wish to express ${\cal Z}_r$ in
terms of the new variables and hence in terms of $x$. From
(\ref{eq:defineH}) we have that
\be
\label{eq:zedroot}
{\cal Z}_r = \frac{1}{2x} \left[ 1 - \epsilon x^2 - \frac{f}{H} \right]
\ee
and any non-analyticity in ${\cal Z}_r$ is due to the singular
behaviour of $f/H$, which is given by
\be
\label{eq:fhinkt}
\frac{f}{H} = \frac{k}{1-2 \tau}.
\ee
Since the term in the square brackets is a function of $x^2$ it will
be convenient to define a variable $q=x^2$ (bear in mind that we have
already defined $f=x^{2/3}$, so that $q=f^3=x^2$).
Given the definition of $k$,
\be
k^3 = f^3 \frac{H^\frac{3}{2}}{\tau^\frac{3}{2}} = \frac{q}{\tau} (1-2
\tau),
\ee
so that
\be
\label{eq:tau}
\tau = \frac{q}{k^3 +2 q}
\ee
and substituting this into (\ref{eq:tauquad}) yields
\be
\left(\frac{q}{k^3}\right)^2 (A+2B+4C) + \left(\frac{q}{k^3}\right)
(B+4C) + C =0 .
\ee
Using equations (\ref{eq:funca}) to (\ref{eq:funcc}) this gives
\be
\label{eq:kquartic}
k^4 - 2 \epsilon q k^3 - k^2 \left[ 3 \epsilon^2 q^2 - 4 q (\eta +
\epsilon) +1 \right] + 8 q k - 4 \epsilon q^2 =0 ,
\ee
which is quartic in $k$, so that we can in principle solve for $k(q)$.
This leaves the question of which root should be chosen.
In the limit $q \to 0$, we have from (\ref{eq:zedroot}) that $f/H \to 1$,
$H \to 0$ and $\tau \to 0$. Thus from (\ref{eq:fhinkt}) we require $k
\to 1$, looking at (\ref{eq:kquartic}) we see that at $q=0$ it becomes
$k^2 \left(k^2-1 \right)=0$ and the root we want is the one for which
$k(q=0)=1$.

Using (\ref{eq:fhinkt}) and (\ref{eq:tau}) we can express $f/H$ in
terms of $k$ and $q$.
Thus finally we have a formula for the grand canonical partition function,
\be
\label{eq:finalsolution}
{\cal Z}_r= \frac{1}{2x} \left[ 1 - \epsilon x^2 - k - \frac{2 x^2}{k^2}
\right],
\ee
where $k(x^2)$ is a calculable function given by solving (\ref{eq:kquartic}).

At this point the simplest course of action was to write a computer
program, which could evaluate (\ref{eq:finalsolution}) at different
values of $x$ and, by slowly increasing $x$ from zero, locate the
closest singularity to the origin.
The non-analyticity occurs when
$k(x^2)$, which is real for small real $x$, becomes complex as $x$ is
increased.

\begin{figure}[phtb]
\caption{Plot of (a) $\frac{\partial^3 \mu_c}{\partial (p t)^3}$;
 (b) $\frac{\partial^3 \mu_c}{\partial (p t)^2 \partial t}$ for a
range of $p t$ at $t=1$, taking $\mu_c=\mu_c(pt,t)$.}
\label{fig:thirdd}
\begin{picture}(120,90)(0,0)
\centerline{\psfig{figure=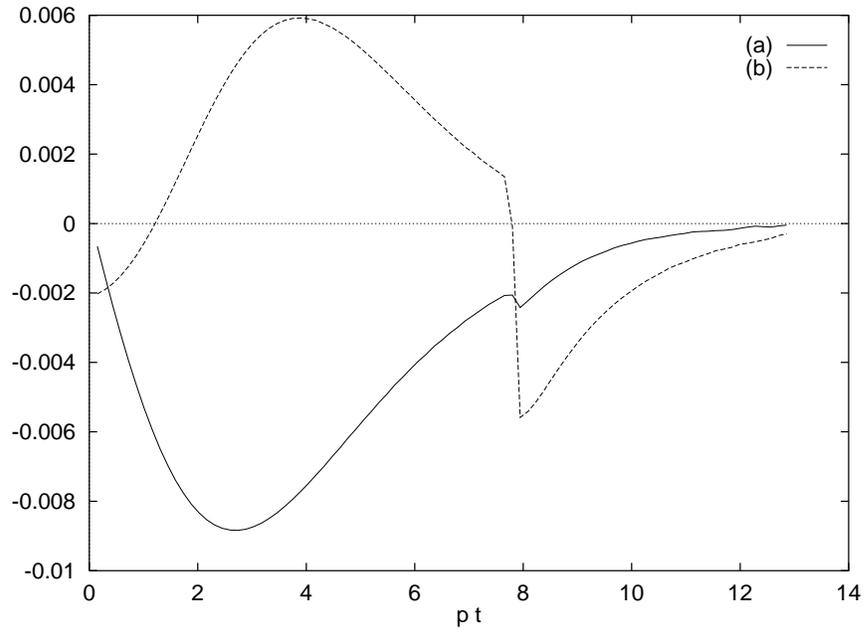,width=12 cm}}
\end{picture}
\end{figure}

\begin{figure}[phtb]
\caption{Plot of $(1- 6 \tau_c)$ for the two-term model}
\label{fig:reftau}
\begin{picture}(100,100)(0,0)
\centerline{\psfig{figure=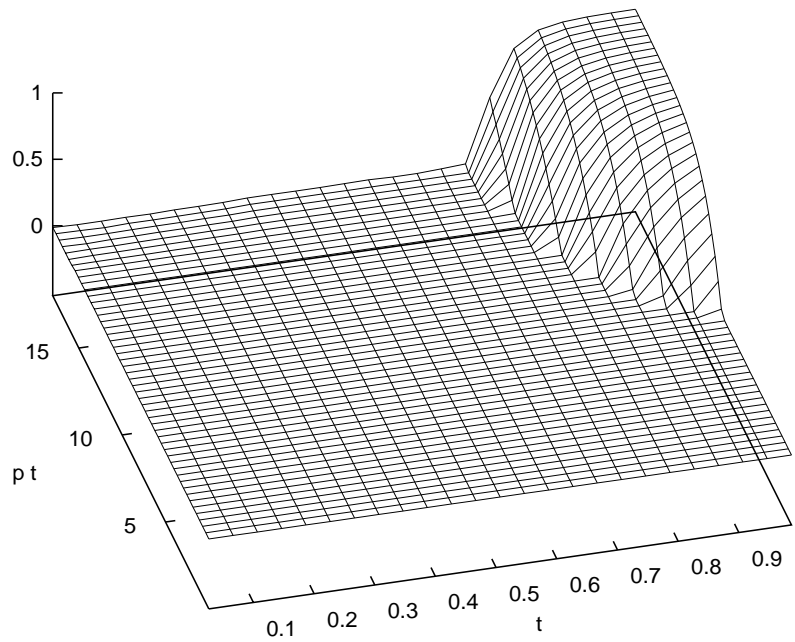,width=15 cm}}
\end{picture}
\end{figure}

Examination of the data for $\mu_c$ indicates that there is a third
order phase transition (see fig~\ref{fig:thirdd}, which plots various third
derivatives of $\mu_c$). The location of the transition
is most clearly seen in fig~\ref{fig:reftau},
which plots the quantity $1-6 \tau$
evaluated at $x_c$; the axes used are $t$ and the product $p t$. There
are two regions one for which $\tau_c=1/6$ and the other for which
$ 0 <\tau_c < 1/6$.

\section{The critical line}
In this section we are going to derive an equation for the critical
line in the two-term model and determine the exponent $\gst$ for the
different regions of the phase diagram.

It will be convenient to rewrite (\ref{eq:kquartic}) as
\be
\label{eq:kfour}
k^4 + a_3 k^3 + a_2 k^2 + a_1 k + a_0 =0 ,
\ee
where we have defined the coefficients $a_0, \cdots, a_3$. Away from
the critical line the closest singularity to the origin is due to a
double root in this equation and thus, evaluating everything at $x_c$,
\be
\label{eq:kthree}
4 k^3 + 3 a_3 k^2 + 2 a_2 k + a_1 =0.
\ee
On the critical line we have a triple root and hence
\be
\label{eq:ktwo}
6 k^2 + 3 a_3 k +a_2 =0 .
\ee
Now we have three equations in four unknowns ($k, q, \epsilon, \eta$)
giving a one parameter solution. In general it would be difficult to
solve these equations, but in this case there is a short cut; we know
that on the critical line $\tau_c=1/6$ and so $4 q = k^3$ (from
(\ref{eq:tau})).
Substituting this into (\ref{eq:kthree}) to eliminate $a_1$ yields
\be
6 k^2 + 3 a_3 k + 2 a_2 =0.
\ee
Comparing this with (\ref{eq:ktwo}) we have $a_2=0$ and $2k+a_3=0$.
Introducing a parameter $e$ gives
\be
\epsilon  = \frac{4}{e^2}, \mgap
k=e, \mgap
(\eta + \epsilon) = \frac{1}{e^3} \left(1+3 e^2 \right), \mgap
q=\frac{1}{4} e^3 .
\ee
It is easy to check that (\ref{eq:kfour}) is satisfied
and to derive a formula for
the critical line,
\be
\eta = \frac{\sqrt{\epsilon}}{2} \left( \frac{\epsilon}{4} +3 \right)
- \epsilon .
\ee
At $t=1$, $\epsilon = \eta= 2^{p} -1 $;
the above equation gives $\sqrt{\epsilon}=2(4+\sqrt{13})$ and hence
the critical line intercepts the $t=1$ axis at $p \approx 7.860327$.
For $p \to \infty$, $\eta \sim \epsilon^{3/2}$ and the intercept with
the line $p=\infty$ occurs at the root of
\be
(1+t) = \left(1+{t}^2 \right)^{\frac{3}{2}},
\ee
which is $t \approx 0.6124088$ (note that $t=0$ is not the correct
root).

In order to calculate $\gst$ it is necessary to find out how the
derivatives of ${\cal Z}_r$ diverge as $q \to q_c$. Putting
$\overline{\cal Z} \equiv 2 x {\cal Z}_r$,
\be
\label{eq:dzdq}
\frac{\partial \overline{\cal Z}}{\partial q} = - \epsilon -
\frac{2}{k^2} - \frac{1}{q'} \left[ 1 - \frac{4 q}{k^3} \right],
\ee
where $q' \equiv \frac{\partial q}{\partial k}$; note that $q'(q_c)=0$
and $q''(q_c) \neq 0$ in general, but that on the critical line
$q''(q_c)=0$.
For the region at large $p$ and $t$, $\tau_c \neq 1/6$ and hence
\be
\frac{\partial \overline{\cal Z}}{\partial q} \sim \frac{1}{q'} \sim
\frac{1}{\sqrt{q-q_c}} .
\ee
Noting that for rooted graphs
\be
\frac{\partial \overline{\cal Z}}{\partial q} \sim (q-q_c)^{- \gst},
\ee
we have $\gst=1/2$ (that is, this is a tree-like region).
For the region at small $p$ and $t$, $\tau_c=1/6$ and the first derivative of
$\overline{\cal Z}$ is finite. Differentiating (\ref{eq:dzdq}) with
respect to $q$ we find that
\be
\frac{\partial^2 \overline{\cal Z}}{\partial q^2} \sim
\frac{1}{\sqrt{q-q_c}} \sim (q-q_c)^{- (\gst +1)}
\ee
and hence $\gst =-1/2$ (the pure gravity value). On the critical line
\be
\frac{\partial \overline{\cal Z}}{\partial q} \sim
\frac{1}{(q-q_c)^{\third}},
\ee
giving $\gst=1/3$. Thus the third order phase transition in the
two-term model is a transition from an unmagnetized non-tree-like
region to a tree-like phase. This provides good evidence for the
existence of such a transition in the full model. Presumably as more
terms are included in the free energy, the critical line will move
somewhat and it is not entirely obvious whether the intercept with the
line $p=\infty$ will occur at non-zero $t$ or $t=0$ in the full model.

The original aim of solving this model was to determine the extent of
the tree-like region, in particular we are interested in finding the
point at which the tree-like, magnetized and unmagnetized phases meet
(assuming that such a point exists); the value of $p$ at this point is
denoted $p^*$. In previous work we made an estimate of the
location of the tree-like region based on the region for which $\mu_c$
was approximately linear. It will be useful to compare the location of
the critical line with the boundary of the approximately linear region
(referred to as the ``knee'' of the solution); the linear region can be
defined to be that for which $\left\vert
\frac{\partial \mu_c}{\partial (pt)} \right\vert < \delta$. We
will take $\delta=0.0168$ giving $p t \approx 6 + 2.7 t + O(t^2)$ at
the knee; this corresponds to our previous estimate that the graphs
were tree-like for $p t \gtsim 6$ at small $t$~\cite{paper2}.

\begin{figure}[tbhp]
\caption{Phase diagram for the two-term model: (a)~Phase
line (b)~Knee ($\delta=0.0168$) (c) The line $t=0.73$}
\label{fig:phaseknee}
\begin{picture}(150,110)(0,0)
\centerline{\psfig{figure=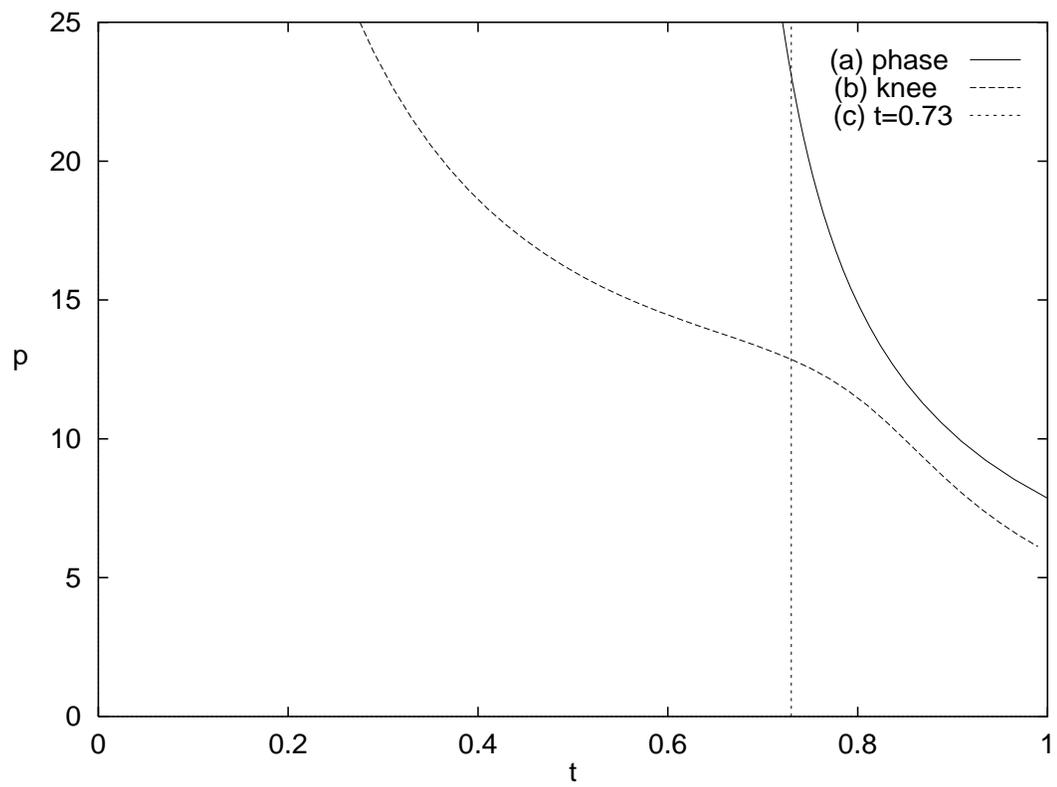,width=15 cm}}
\end{picture}
\end{figure}

In fig~\ref{fig:phaseknee} is plotted the line of the phase transition
in the $p$--$t$ plane and also the location of the knee.
The tree-like phase is inside the linear region and the two
lines are quite close in the area of interest (near $t \approx
0.73$). Using the phase line as an approximation to the location of
the phase transition in the full model we can estimate the point at
which it intercepts the boundary of the magnetized region. Although we
do not know exactly where the critical line separating the magnetized
and unmagnetized phases lies, we have that $t_c \approx 0.73$ for the
single spin Ising model~\cite{BouKaz1} and we know from Monte Carlo
simulations~\cite{BaiJoh2,ADJT,BFHM} that $t_c$ only increases slowly with $p$.
Taking $t=0.73$ gives an estimate of $p^* \sim 23$,
but this is likely to be an over-estimate since the phase line changes
rapidly with $t$ and the value of $t$ that we are using is too small
(it corresponds to the critical coupling for $p=1$). An estimate using
our old criterion based on the knee would give a value of $p^* \sim 13$.
Overall then we conclude that $p^* \sim 13$ -- $23$ and that the
corresponding value of the central charge,
above which there exists a tree-like phase in the full model,
is $c \sim 6$ -- $12$.

\section{Conclusion}

\begin{figure}[bth]
\caption{Possible phase diagrams for model I}
\label{fig:phase}
\begin{picture}(140,120)(0,0)
\centerline{\psfig{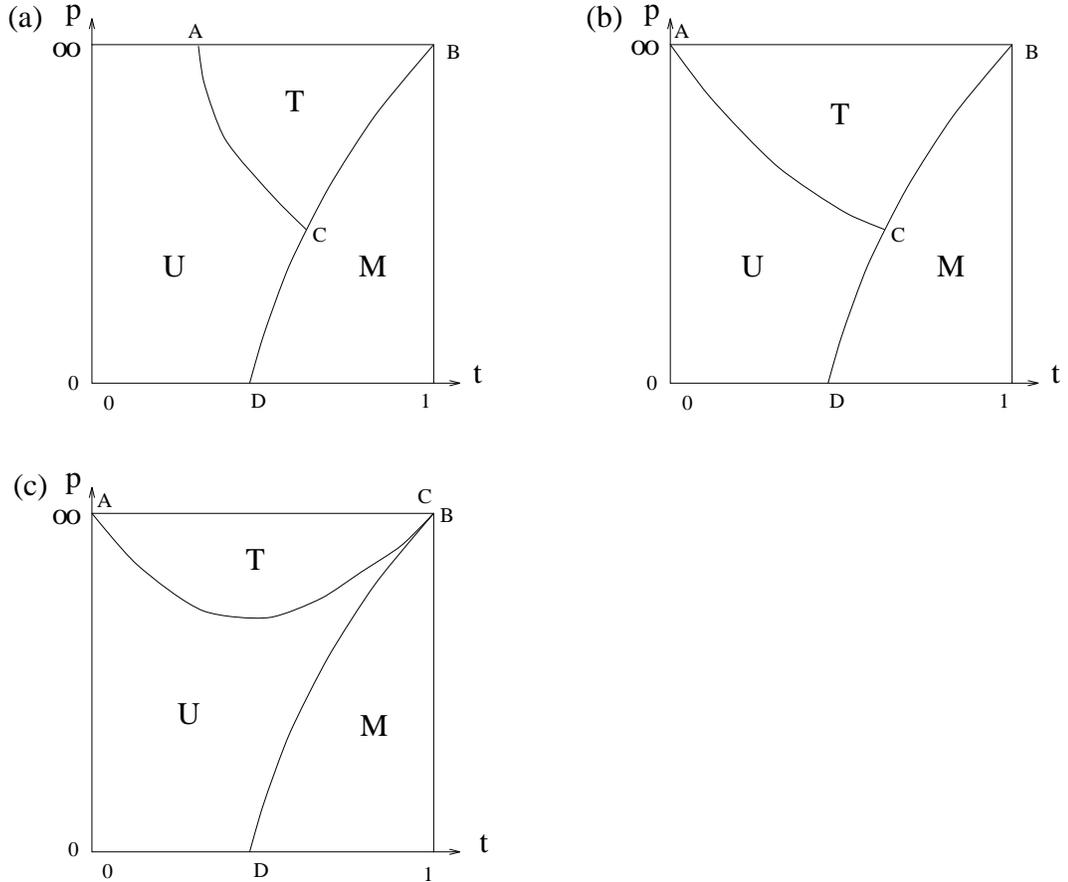}}
\end{picture}
\end{figure}

In fig~\ref{fig:phase} we have drawn some possible forms for the phase
diagram of model~I. The magnetized region is labelled M, the
tree-like region T and the remaining unmagnetized non-tree-like region U.
If the critical line in the two-term model were an
accurate representation of that in the full model then the correct phase
diagram would be fig~\ref{fig:phase}a. However, one expects that as more
terms are included in the truncated model the location of the
critical line will change and it is not entirely clear where point A
will move to. It is possible that A tends towards the $t=0$ axis as
more terms are added in which case fig~\ref{fig:phase}b would be the
correct diagram. Other possibilities, such as fig~\ref{fig:phase}c
or the absence of a tree-like region, can not be entirely ruled out,
but seem unlikely given the results presented in this letter and those
due to Wexler~\cite{Wex1,Wex2}.
The two-term model has a third order transition with
$\gst=\third$ on the critical line AC and the full model may well show
the same behaviour. It is interesting to note that the model in
reference~\cite{JonWhe94} also has $\gst=\third$ on the line
separating the branched polymer and unmagnetized phases (in this model
$\gst=\frac{1}{4}$ at the point where the three phases meet).
The order of magnitude estimate of $p^* \sim 13$--$23$ (point C in
fig~\ref{fig:phase}) is similar to the
largest value of $p$ (ie $p=16$) used in simulations so far. It is
to be hoped that future simulations will manage to locate the onset of
the branched polymer phase. One should note that our estimate of $p^*$
is not sufficiently accurate to exclude the possibility that $p^*=2$,
although numerical simulations seem to rule this out.
In conclusion then, significant progress has been made in elucidating
the nature of the phase diagram for the multiple Ising model on
dynamical planar \ph3 graphs, in
particular in the large $p$ limit, but much work remains to be done
before the nature of the $c=1$ barrier is fully understood.

\section*{Acknowledgment}

I would like to thank John Wheater for useful discussions concerning
this work and to acknowledge the support of PPARC through the research
studentship GR/H01243.

%%%%%%%%%%%%%%%%%%%%%%%%%%%%%%%%%%%%%%%%%%%%%%%%%%%%%%%%%%%%%%%%%%%%%
%% The Bibliography
%%%%%%%%%%%%%%%%%%%%%%%%%%%%%%%%%%%%%%%%%%%%%%%%%%%%%%%%%%%%%%%%%%%%%

\end{document}